\documentclass[copyright,creativecommons]{eptcs}
\providecommand{\event}{PLACES 2017} 
\usepackage{breakurl}             
\usepackage{underscore}           

\usepackage{tikz-qtree}
\usetikzlibrary{trees}

\tikzset{
  treenode/.style = {align=center, inner sep=1pt, text centered,
    font=\sffamily},
  invisible/.style={opacity=0.5},
}    

\usepackage{proof}
\usepackage{latexsym}
\usepackage{amsmath}
\usepackage{mathtools}

\usepackage{titlesec}

\titlespacing*{\section}
{0pt}{10pt}{10pt}
\titlespacing*{\subsection}
{0pt}{10pt}{10pt}
\titlespacing*{\paragraph}
{0pt}{10pt}{10pt}

\newcommand{\ap}{\mathbin{+\!\!\!+}}

\newcommand{\main}{\circ}
\newcommand{\shdw}{\bullet}
\newcommand{\shadows}{\mathit{shadows}}
\newcommand{\I}[1]{\mathbin{I_{#1}}}
\newcommand{\Isame}{\mathbin{I^{\mathit{same}}}}
\newcommand{\Idiff}[1]{\mathbin{I^{\mathit{diff}}_{#1}}}

\newcommand{\seq}{\, \vdash \, }

\title{Generating Representative Executions \\{} \large{\normalfont{Extended Abstract}}}
\author{Hendrik Maarand \quad\quad Tarmo Uustalu
\institute{Dept.\ of Software Science, Tallinn University of Technology}}
\def\titlerunning{Generating Representative Executions}
\def\authorrunning{H. Maarand \& T. Uustalu}
\begin{document}
\maketitle

\begin{abstract}
  Analyzing the behaviour of a concurrent program is made difficult by the
  number of possible executions. This problem can be alleviated by applying the
  theory of Mazurkiewicz traces to focus only on the canonical representatives
  of the equivalence classes of the possible executions of the program. This
  paper presents a generic framework that allows to specify the possible behaviours of
  the execution environment, and generate all Foata-normal executions of a program,
  for that environment, by discarding abnormal executions
  during the generation phase. The key ingredient of Mazurkiewicz trace theory,
  the dependency relation, is used in the framework in two
  roles: first, as part of the specification of which executions are
  allowed at all, and then as part of the normality checking algorithm, which is
  used to discard the abnormal executions. The framework is
  instantiated to the relaxed memory models of the SPARC hierarchy.
\end{abstract}
 
\section{Introduction} \label{introduction}
Let us consider a fragment from Dekker's mutual exclusion algorithm as an
example.
\begin{center}
    \begin{tabular}{l | l}
      \multicolumn{2}{c}{Init: \texttt{x = 0; y = 0;}} \\                
      \hline                         
      \multicolumn{1}{c |}{$P_1$} & \multicolumn{1}{| c}{$P_2$} \\
      \hline
      (a) \texttt{[x]  := 1}   & (c) \texttt{[y]  := 1}   \\             
      (b) \texttt{r1 \ := [y]} & (d) \texttt{r2 \ := [x]} \\             
      \hline                                                             
      \multicolumn{2}{c}{Observed? \texttt{r1 = 0; r2 = 0;}}             
    \end{tabular}                                                        
\end{center}
This is a concurrent program for two processors, $P_1$ and $P_2$, where
$x$ is the flag variable for $P_1$ that is used to communicate that $P_1$ wants
to enter the critical section and $y$ is for $P_2$. A processor may enter the
critical section, if it has notified the other processor by setting its flag
variable to $1$, reading the flag variable of the other processor and checking that
it is $0$. We are interested in whether it is possible, starting from an initial
state where both $x$ and $y$ are $0$, that both processors see each others' flag
variables as $0$, meaning that both processors enter the critical section. Here we
are interested in the mutual exclusion property, that at most one processor can
enter the critical section.

In the interleaving semantics of Sequential Consistency (SC), the
above program can have the following executions: \textit{abcd},
\textit{cdab}, \textit{acbd}, \textit{cabd}, \textit{acdb},
\textit{cadb}. Out of these six, the four last executions are actually
equivalent (in the sense that from the same initial state they will
reach the same final state) and for our purposes it is enough to check
the final state of only one of them. We can observe that the mutual
exclusion property is satisfied. The situation is different, if we
consider the possible executions on a real-world processor, like x86,
which follows the Total Store Order (TSO) model~\cite{x86-tso}. Under
TSO, it is possible for writes to be reordered with later reads from
the same processor, resulting in an execution that is observable as
\textit{bdac}. This does not satisfy the mutual exclusion property.

In this paper, we seek to alleviate the difficulty analyzing the large
numbers of executions concurrent programs, especially on relaxed
memories, generate, by applying the theory of Mazurkiewicz traces to
focus only on some type of canonical representatives of the
equivalence classes of the possible executions of the program. We
present a generic framework for interpreting concurrent programs under
different semantics, so that only executions in the Foata normal form
(corresponding to maximal parallelism) are generated. We instantiate
the framework to the relaxed memory models of the SPARC
hierarchy. This work is in the vein of partial order reduction
techniques for analysis of systems, which are widely used especially in model checking and have also been applied to relaxed memories, e.g., by Zhang et al.~\cite{zhang}. The novelties here are that the different memory models
are modelled uniformly based on a flexible notion of a backlog of
shadow events, using a standard normal form from trace theory, and
using generalized traces (with a dynamic independency relation) to be
able to define execution equivalence more finely, resulting in bigger
and fewer equivalence classes. The framework has been prototyped in
Haskell where one can easily separate the phases of generating the
tree of symbolic executions of a program, discarding the abnormal
executions, and running the tree of symbolic executions from an
initial state. This separation can be made without a performance
penalty thanks to lazy evaluation.


\section{Mazurkiewicz Traces} \label{traces}
An execution (or a run) of a sequential program can be represented as
a sequence of symbols that record the events caused by the program in
the order that they occurred. Such a sequence is a string over some
(finite) alphabet $\Sigma$. An execution of a concurrent program can
be represented as an interleaving of the executions on the processors
involved, thereby reducing concurrency to non-deterministic
choice. Mazurkiewicz traces~\cite{mazurkiewicz} (or just traces) are a
generalization of strings, where some of the letters in the string are
allowed to commute. This allows representation of non-sequential
behaviour. In other words, traces are equivalence classes of strings
with respect to a congruence relation that allows to commute certain
pairs of letters.

A dependency relation $D \subseteq \Sigma \times \Sigma$ is a reflexive and
symmetric binary relation. $a \ D \ b$ if and only if the events $a$ and $b$
can be causally related, meaning that the two events cannot happen
concurrently. The complement of the dependency relation,
$I = (\Sigma \times \Sigma) \setminus D$, is called the independency
relation. If $a \ I \ b$, then the strings \textit{sabt} and \textit{sbat}
represent the same non-sequential behaviour. Two strings $s, t \in \Sigma^*$ are
said to be Mazurkiewicz equivalent, $s \equiv_D t$, if and only if $s$ can be
transformed to $t$ by a finite number of exchanges of adjacent, independent
events. For example, if $\Sigma = \{ a, b, c, d \}$ and $a \ I \ c$ and
$b \ I \ d$ then the trace \textit{acbd} represents the strings \textit{acbd},
\textit{cabd}, \textit{acdb} and \textit{cadb}.

For our purposes, standard Mazurkiewicz traces are not enough and
therefore we turn to the generalized Mazurkiewicz traces of Sassone et
al.~\cite{sassone}. In generalized Mazurkiewicz traces, the dependency
relation is dynamic, it depends on the current context, which is the
partial execution that has been performed so far. The dependency
relation for a prefix $s$ will be denoted by $D_s$ and the subscript
is omitted, if the relation is static. Besides $D_s$ having to be
reflexive and symmetric for any $s$, $D$ must satisfy some sanity
conditions. Most importantly, if $s\equiv_D t$, then it must be the
case that $D_s = D_t$. In this setting, the strings \textit{sabt} and
\textit{sbat} are considered equivalent, if $a \ I_{s} \ b$.

\paragraph{Normal Forms} \label{foata}
As traces are equivalence classes, it is reasonable to ask what the
canonical representative or normal form of a trace is. There are two well-known
normal forms for traces, the lexicographic and Foata~\cite{foata} normal forms. We
are going to look at Foata normal forms for our purposes.

A step is a subset $s \subseteq \Sigma$ of pairwise independent
letters.  The Foata normal form of a trace is a sequence $s_1 \dots
s_k$ of steps such that the individual steps $s_1,\dots,s_k$ are
chosen from the left to the right with maximal cardinality. Since each
step consists of independent letters, a step can be executed in
parallel, meaning that the Foata normal form encodes a maximal
parallel execution. For example, if $\Sigma = \{ a, b, c, d \}$ and $a
\ I \ c$ and $b \ I \ d$, then the Foata normal form of \textit{acbd}
is \textit{(ac)(bd)}.

We are interested in checking whether a given string is in normal form
according to a given dependency relation. As a convenience, we also
assume to have an ordering $\prec$ on $\Sigma$ that is total on events that
are independent. A string is in Foata normal form, if it can be split
into a sequence of steps $s_1,\dots,s_k$ so that concatenation of
the steps gives the original string and the following conditions are
satisfied:
\begin{enumerate}
\item for every $a, b \in s_i$, if $a \neq b$ then $a \ I_i \ b$;
\item for every $b \in s_{i+1}$, there is an $a \in s_i$ such that $a \ D_i \ b$;
\item for every step $s_i$, the letters in it are in increasing order wrt.\ $\prec$.
\end{enumerate} 
In these definitions, we consider $D_i$ to be the dependency relation
for the context $s_0 \dots s_{i-1}$ and similarly for $I_i$. The first
condition ensures that the events in a step can be executed in
parallel. The second condition ensures that every event appears in the
earliest possible step, i.e., maximal parallelism. The third condition
picks a permutation of a step as a representative of the step. Notice
that if a string is not in normal form, then neither is any string with that
string as a prefix in normal form. This means that when
checking a string for normality by scanning it from the left to the
right, we can discard it as soon as we discover an abnormal prefix.

\section{Framework} \label{framework}

We now proceed to describing our framework for generating
representative executions of a program and its instantiations to
different memory models.

We are going to look at programs executing on a machine that consists
of processors and a shared memory. Each processor also has access to a
local memory (registers). The executions that we investigate are
symbolic, in the sense that we do not look at the actual values
propagating in the memory, but just the abstract actions being
performed. Still, our goal is to find the possible final states of a
program from a given initial state. The idea is that once the symbolic
executions have been computed, the canonical executions can be picked
and the final state needs to be computed only for those. This can
be done lazily, meaning that the evaluation of a particular execution for the given 
initial state is cancelled immediately, if it is discovered that the
execution is not normal. 

The language for our system consists of arithmetic and boolean expressions and
commands. An arithmetic expression is either a numeral value, a register, or an
arithmetic operation. A boolean expression is either boolean constant, a boolean
operation, or a comparison of arithmetic expressions. Commands consist of
assignments to registers, loads and stores to shared memory, and
\textit{if} and \textit{while} constructs.

Our framework is defined on top of the events generated by the
system. We think of events as occurrences of (the phases of) the
actions that executing the program can trigger. An event can be
thought of as a record \textit{(pid, eid, kind, act)} where \textit{pid}
is the identifier of the processor that generated the event,
\textit{eid} is the processor-local identifier of the event, \textit{kind} defines
whether it is a main or a shadow event, and \textit{act} is the action
performed in this event. 
An action can be an operation between registers, a
load from or a store to a variable, or an assertion on registers. An
assertion is used to record a decision made in the unfolding of a
control structure of the program, for example, that a particular
execution is one where the \textit{true} branch of a conditional was
taken. If an assertion fails when an execution is evaluated from a
given initial state, then this execution is not valid for that
initial state.

Since we are interested in modelling different memory models, our
framework is parameterized by an architecture, which characterizes the
behavioural aspects of the system. An architecture consists of four
components. A predicate \textit{shadows} describes whether an action
is executed in a single stage or two stages, generating just one
(main) event or two events (a main and a shadow event).  An
irreflexive-antisymmetric relation \textit{sameDep} describes which
events from a processor must happen before which other events from the
same processor: it plays a role in determining the possible next
events from this processor, but also defines which events from it are
dependent. A relation \textit{diffDep} describes when two events from
different processors are dependent. Finally, a relation $\prec$ orders
independent events. The relations \textit{sameDep} (its
reflexive-symmetric closure) and \textit{diffDep} together determine
the dependency relation in the sense of Mazurkiewicz traces and
$\prec$ is the relation used to totally order the events within a
step.

In the previous paragraph, we mentioned shadow events. These are the
key ingredients of this framework for modelling more intricate
behaviours, for example, when some actions are non-atomic and this
fact needs to be reflected in the executions by two events, a main
event and a shadow event. TSO, for example, can be described as a
model where writes to memory first enter the processor's write-buffer
and are later flushed from the write-buffer to memory. We consider the
write to buffer to be the main event of the write action and the flush
event to be the shadow event of the write action. Of these two events,
the shadow event is globally observable.

\paragraph{Generating Normal Forms}
The process of generating normal-form executions of a program can be
divided into two stages: lazily generating all executions of the
program and then discarding those not in normal form. 

The executions are generated as follows: if all processors have
completed, then we have a complete execution and we are done,
otherwise we pick a processor that has not yet completed and allow it
to make a small step, then repeat the process. The local configuration
of a processor consists of its residual program, backlog, and the value of a
counter to provide identifiers for the generated events. The small
step 
can either
correspond to beginning the action of the next instruction according
to the program---in which case a new main event is generated and added
to the execution---, or to completing an already started action---in
this case, a shadow event is removed from the processor's backlog and
added to the execution. If the step is to start a new action, then the
\textit{shadows} predicate is used to check whether a new shadow event
should be added to the backlog (if not, the action is completed by the
main event). A side-condition for adding a new main event is that
there are no shadow events in the backlog that are dependent with
it. An event can be removed from the backlog, if it is independent
(according to \textit{sameDep}) of all of the older events in the
backlog. Conditionals like \textit{if} and \textit{while} are expanded
to a choice between two programs, where the choices correspond to the
branches of the conditional together with an assertion of the
condition. The generation of executions is described by the small step
rules in Appendix~\ref{rules}.

The second stage of the procedure is to single out the normal forms
among the generated executions. This is done by checking the normality
of the executions according to the three conditions given in
Section~\ref{foata} for Foata normal forms. The rules for checking the
normality of an execution by scanning it from the left to the right
are given in Appendix~\ref{rules}.

Instead of generating a flat set of executions in the first stage, we
actually generate a tree of executions, so that the prefixes of
executions are shared. Since the process of selecting the canonical
executions (more precisely, discarding the non-canonical ones)
according to the conditions of Foata normal forms can be fused into the
generation stage, we can discard a whole set of executions when we
discover that the current path down the tree violates the normality
conditions. More precisely, walking down the tree, we keep track of
the current prefix (which must be in normal form) and at each node we
check whether the event associated with the node would violate the
normality conditions when added to the prefix. Only if the normality
condition is not violated does the subtree starting from that node
need to be computed actually. 

We require \textit{sameDep a b} to hold at least when \textit{a} and
\textit{b} are main events and $\mathit{eid}\ a < \mathit{eid}\, b$ or
when they are a main event and its shadow event (in which case they
have the same \textit{eid}). We also require that \textit{sameDep a b}
can only hold when $\mathit{eid}\ a < \mathit{eid}\, b$ or when
$\mathit{eid}\ a = \mathit{eid}\, b$ and \textit{a} is a main event
and \textit{b} the corresponding shadow event. Under these
assumptions, we can prove that the total set of executions captured in
the generated tree is closed under equivalence. As the normality
checking stage keeps all normal forms and discards all non-normal
forms, it follows that the pruned set of executions contains exactly
one representative for every execution of the program.

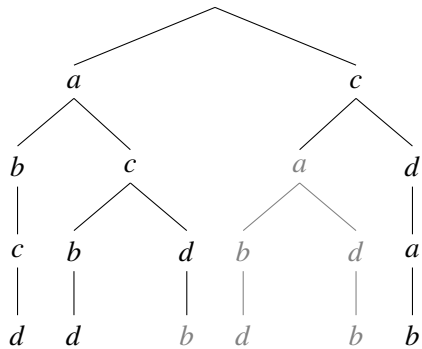
\begin{figure}[h]
  \begin{center}
    \begin{tikzpicture}
      [auto, 
       level 1/.style={sibling distance = 5cm},
       level 2/.style={sibling distance = 2cm}, scale=0.75]
      \node {} 
      child{ node {$a$} 
        child{ node {$b$} 
          child{ node {$c$}
            child{ node {$d$} }  
          } 
        }
        child{ node {$c$} 
          child { node {$b$}
            child { node {$d$} } 
          } 
          child { node {$d$} 
            child { node[invisible] {$b$} } 
          } 
        }
      } 
      child{ node {$c$} 
        child { node[invisible] {$a$} 
          child[invisible] { node {$b$} 
            child { node {$d$} }
          } 
          child[invisible] { node {$d$} 
            child { node {$b$} }
          } 
        } 
        child { node {$d$} 
          child { node {$a$} 
            child { node {$b$} }
          } 
        } 
      };
    \end{tikzpicture}
  \end{center}
  \caption{SC executions of the example program.}
  \label{fig:sc-executions}
\end{figure}

In the introduction, we noted that our example program has six executions under
interleaving semantics, of which four are equivalent. The executions are
depicted in Figure~\ref{fig:sc-executions} and the four equivalent executions
\textit{acbd}, \textit{acdb}, \textit{cabd} and \textit{cadb} are the ones in
the middle. For this program we have that $a \ I \ c$ and $b \ I \ d$. Our
framework would only generate \textit{acbd} out of these four, as this
corresponds to the Foata normal form \textit{(ac)(bd)} and the other three would
be discarded. More precisely, \textit{(ac)(d)} is in normal form, but it cannot be extended by \textit{b}, as neither
\textit{(ac)(db)} nor \textit{(ac)(d)(b)} is in normal form: the first one fails due
to condition 3 and the second one fails due to condition 2. The node $b$ of
this path is shaded in the picture to highlight the place where the normality
condition is violated. For \textit{cabd}, we start checking normality from
\textit{(c)}, which is valid, but neither \textit{(ca)} nor \textit{(c)(a)} is
in normal form and we can discard all executions that start with \textit{ca},
which includes both \textit{cabd} and \textit{cadb}. The subtree at node $a$ is
shaded to highlight this fact.

\section{Instantiation to Relaxed Memory Models}

\paragraph{Sequential Consistency}
In the Sequential Consistency (SC) model~\cite{lamport}, any execution
of a concurrent program is an interleaving of the program order executions of
its component threads. SC can be specified as an architecture in the following
way:
\begin{align*}
\mathit{shadows}         \ a     & = \mathit{false} \\
\mathit{sameDep}         \ a \ b & = \mathit{eid} \ a <  \mathit{eid} \ b  \\
\mathit{diffDep} \ x \ y \ a \ b & = \mathit{crxw} \ a \ b \\
a \prec b & = \mathit{pid} \ a < \mathit{pid} \ b  
\end{align*}

\noindent \textit{crxw a b} represents the
concurrent-read-exclusive-write property, which returns \textit{true},
if events $a$ and $b$ access the same location and at least one of
them is a write. \textit{diffDep} also takes two arguments that are
ignored here, which represent the backlogs of the two processors from
which the events $a$ and $b$ originate from. This information can be
recovered from the prefix of the execution and it is as much
information as we need about the prefix of the execution in the memory
models we consider. We could also just take the prefix of the
execution itself and compute the necessary information. Setting
\textit{shadows} to be always \textit{false} means that all
instructions execute atomically. Setting \textit{sameDep a b} to
require $\mathit{eid} \ a < \mathit{eid} \ b $ means that the events
from the same processor must be generated in program order and cannot
be reordered, which reflects the definition of SC.

\paragraph{Total Store Order}
In the Total Store Order (TSO) model~\cite{sparc1994}, it is possible for a write
action to be reordered with later reads, meaning that writes happen
asynchronously, but at the same time the order of write actions is
preserved. TSO can be specified in the following way:
\begin{align*}
\mathit{shadows}         \ a     & = \mathit{isWrite} \ a \\
\mathit{sameDep}         \ a \ b & = 
\mathit{isMain}\ a \ \land \ \mathit{isMain} \ b \ \land \ \mathit{eid}\, a < \mathit{eid}\ b \\
         &  \lor \ \mathit{isMain} \ a \ \land \ \mathit{isShadow} \ b
           \ \land \ \mathit{eid}\, a == \mathit{eid}\ b  \\
                                     & \lor \ \mathit{isShadow} \ a \
                                       \land \
                                   \mathit{isShadow} \ b \ \land \ \mathit{eid}\, a < \mathit{eid}\ b \\
\mathit{diffDep} \ x \ y \ a \ b & = \mathit{crxw'} \ x \ y \ a \ b \\
                       a \prec b & = \mathit{pid} \ a < \mathit{pid} \
                                   b \ \lor \ \mathit{pid} \ a ==
                                   \mathit{pid} \ b \ \land \ \mathit{eid}
                                   \ a < \mathit{eid} \ b 
\end{align*}
\textit{crxw'} is like \textit{crxw}, except that it considers shadow
write events instead of main write events as the global write events,
and read events as global only if they access the memory. This is where we need
generalized Mazurkiewicz traces, since if there is a pending write to the
location of the read, then the read action would not read its value from memory
and thus could not be dependent with events from other
processors. 

We consider the main event of a write instruction to be the write to buffer and
the shadow event to be the flushing of the write from buffer to memory. TSO can
be thought of as a model where every processor has a shadow processor and all
events on every main processor are in program order, all of the events on the
associated shadow processor are in program order and an event on the shadow processor must
happen after the corresponding event on the main processor.  Our example from
introduction has the following traces in Foata normal form under TSO:
\textit{(ac)(a'c')(bd)}, \textit{(ac)(a'b)(c'd)}, \textit{(ac)(c'd)(a'b)} and
\textit{(ac)(bd)(a'c')} where \textit{a'} stands for the shadow event of
\textit{a}. The last of these is the one rejected by SC.

\paragraph{Partial Store Order}

The Partial Store Order (PSO) model~\cite{sparc1994} allows the
reorderings of TSO, but it is also possible for a write to be
reordered with a later write to a different location. This can be
thought of as having a separate write buffer for every variable. PSO can be
specified as TSO with the exception of the \textit{sameDep} relation:
\begin{align*}
\mathit{sameDep} \ a \ b & = 
\mathit{isMain}\ a \ \land \ \mathit{isMain} \ b \ \land \ \mathit{eid}\, a < \mathit{eid}\ b \\
                         &  \lor \ \mathit{isMain} \ a \ \land \ \mathit{isShadow} \ b
           \ \land \ \mathit{eid}\, a == \mathit{eid}\ b  \\
                         & \lor \ \mathit{isShadow} \ a \ \land \
                           \mathit{isShadow} \ b \ \land \
                           \mathit{eid}\, a < \mathit{eid}\ b \ \land \
                           \mathit{var} \ a == \mathit{var} \ b 
\end{align*}

\noindent Intuitively, this corresponds to PSO, since it is like TSO except for
the dependency relation on events from the same processor, where the shadow
events are dependent only if they are to the same location, which allows one to
reorder writes to different locations.

\paragraph{Relaxed Memory Order}
The Relaxed Memory Order~\cite{sparc1994} (RMO) model only enforces program
order on write-write and read-write instruction pairs to the same variable and
on instruction pairs in dependency, where the first instruction is a
read. Dependency on instruction pairs here means that there is data- or
control-dependency between the instructions. We can specify RMO in the following
way:
\begin{align*}
\mathit{shadows} \ a     & = \mathit{true} \\
\mathit{sameDep} \ a \ b & = 
\mathit{isMain}\ a \ \land \ \mathit{isMain} \ b \ \land \ \mathit{eid}\, a < \mathit{eid}\ b \\
                         &  \lor \ \mathit{isMain} \ a \ \land \ \mathit{isShadow} \ b \
           \land \ \mathit{eid}\, a == \mathit{eid}\ b  \\
                         & \lor  \ \mathit{isShadow} \ a \ \land \
                           \mathit{isShadow} \ b \
                           \land \ \mathit{eid} \ a < \mathit{eid} \ b \\
    & \qquad \land \ (\mathit{var} \ a == \mathit{var} \ b \ 
              \land \ (\mathit{isWrite} \ a \ \lor \ \mathit{isRead} \ a) \
              \land \mathit{isWrite} \ b \\  
    & \qquad \qquad      \lor \ \mathit{dataDep} \ a \ b \
                         \lor \ \mathit{controlDep} \ a \ b) \\
\mathit{diffDep} \ x \ y \ a \ b & = \mathit{crxw''} \ x \ y \ a \ b \\
                       a \prec b & = \mathit{pid} \ a < \mathit{pid} \
                                   b \ \lor \ \mathit{pid} \ a ==
                                   \mathit{pid} \ b \ \land \ \mathit{eid}
                                   \ a < \mathit{eid} \ b 
\end{align*}
\textit{crxw''} is like \textit{crxw'} except that it considers shadow reads and
shadow writes 
as the global read and write events. As for TSO and PSO, a shadow read
is considered global, if it actually reads its value from memory, which in this model happens, if there is no older shadow write to the same location in the backlog. We
consider events $a$ and $b$ to be in data-dependency, if $a$ reads a
register that is written by $b$. We consider two events to be in
control-dependency, if the older one is a conditional and the newer one is a write.

\subsection{Fences}
In models like TSO, PSO and RMO that allow the reordering of some events, it
becomes necessary to be able to forbid these reorderings in certain situations,
to rule out relaxed behaviour. Our example from introduction does not behave
correctly on TSO, where it is possible for both processors to read the value 0. To
avoid this situation, it is necessary to make sure that both processors first
perform the write and when the effects of the write operation have become
globally visible they may perform the read. With this restriction the program
behaves correctly on TSO and the way to achieve this is to insert a fence
between the write and read instructions.

In our framework, fences are described by two parameters that can take
the values \textit{store} or \textit{load}, which indicate between
which events the ordering is enforced. Under SC, the fence
instructions can be ignored since no reorderings are possible. To be
able to restore sequentially consistent behaviour, TSO requires
store-load fences, PSO requires also store-store fences, and RMO
requires all four kinds of fences. For TSO, PSO, and RMO, the idea is
that fences have shadow events and the \textit{sameDep} relation is
modified to disallow unwanted reorderings. Our example program
requires a store-load fence, so that the read operations appearing
after the fence cannot be performed before the write operations
appearing before the fence have completed. This means that
\textit{sameDep} must be modified to consider a shadow store-load
fence to be dependent with all older shadow write events and all newer
read events. Dependence with a shadow event prevents the fence
event from being removed from the backlog until the older dependent events
have been removed and it also prevents removing the newer dependent
events until the fence has been removed from the backlog. Likewise, a
new main read event cannot be added to the execution, if there is a
store-load fence event in the backlog.  The idea is similar for the other types
of fences.

\section{Related Work}
Relaxed memory consistency models and their specification and
verification tasks have been an extensive research topic. Owens et
al.~\cite{x86-tso} showed that x86 adheres to TSO model and they gave
both operational and axiomatic models. Alglave~\cite{alglave} defined
a framework in an axiomatic style for working with relaxed memory
models, which is also generic in the sense that different memory
models can be represented by specifying which relations are considered
global. Generating the possible executions in our framework turns out
to be quite similar to an executable specification for RMO given by
Park and Dill~\cite{park}, more precisely, our notion of backlog seems
to correspond to the reordering box used there. Boudol et
al.~\cite{boudol12} defined a generic operational semantics that
captures TSO, PSO and RMO and uses temporary stores that again are
similar to our backlogs; they did not however consider any partial
order reduction of the set of executions of a program. As mentioned
before, due to the interest in exploring the full set of executions by
constructing it explicitly and the use of trace theory, which is the
foundation for partial order reduction \cite{godefroid}, this work is
also close to methods based on model checking, like Zhang et al.'s
\cite{zhang} and Abdulla et al.'s \cite{abdulla}.  An executable
specification was also given by Yang et al.~\cite{yang}. Their approach
is based on axiomatic specifications and an execution is found
by searching for an instantiation that satisfies all of the
constraints, either by Prolog or a SAT solver.

\section{Conclusion}
We have presented a generic framework for finding canonical
representatives of equivalence classes of the possible executions of a
program. The framework proceeds by lazily generating all executions of
the given program and discards all those that are not in Foata normal
form. The framework allows to uniformly represent the semantics of a
certain class of relaxed memory models, which we have illustrated by
encoding the models from the SPARC hierarchy in terms of our
framework. An instantiation of the framework to a particular model
specifies which executions can occur at all for the given program and
which of those are equivalent, i.e., correspond to one generalized
Mazurkiewicz trace, representable by its normal form.

We plan to continue this work by elaborating on the formal aspects of
the framework. We have formalized soundness and completeness of Foata
normalization of (standard) traces in the dependently typed functional
language Agda---any string is equivalent to its normal form, and if a
string is equivalent to a normal form, it is that string's normal
form. This development can be scaled for generalized traces, adapted
to prove that the tree filtering algorithm keeps exactly one
representative of each equivalence class of executions, to then move
on to formalization of specifications of memory models.


\paragraph{Acknowledgments}
This research was supported by the Estonian Ministry of Education and
Research institutional research grant no.~IUT33-13 and the ERDF funded
CoE project EXCITE (2014-2020.4.01.15-0018).


\begin{thebibliography}{10}
\providecommand{\bibitemdeclare}[2]{}
\providecommand{\surnamestart}{}
\providecommand{\surnameend}{}
\providecommand{\urlprefix}{Available at }
\providecommand{\url}[1]{\texttt{#1}}
\providecommand{\href}[2]{\texttt{#2}}
\providecommand{\urlalt}[2]{\href{#1}{#2}}
\providecommand{\doi}[1]{doi:\urlalt{http://dx.doi.org/#1}{#1}}
\providecommand{\bibinfo}[2]{#2}

\bibitemdeclare{inproceedings}{abdulla}
\bibitem{abdulla}
\bibinfo{author}{P.~A. \surnamestart Abdulla\surnameend},
  \bibinfo{author}{S.~\surnamestart Aronis\surnameend}, \bibinfo{author}{M.~F.
  \surnamestart Atig\surnameend}, \bibinfo{author}{B.~\surnamestart
  Jonsson\surnameend}, \bibinfo{author}{C.~\surnamestart
  Leonardsson\surnameend} \& \bibinfo{author}{K.~\surnamestart
  Sagonas\surnameend} (\bibinfo{year}{2015}): \emph{\bibinfo{title}{Stateless
  Model Checking for {TSO} and {PSO}}}.
\newblock In: \bibinfo{editor}{C.~\surnamestart Baier\surnameend} \&
  \bibinfo{editor}{C. \surnamestart Tinelli\surnameend}, editors: {\sl \bibinfo{booktitle}{Proc.\ of 21st Int.\ Conf.\ on Tools and
  Algorithms for the Construction and Analysis of Systems, TACAS~2015}}, {\sl
  \bibinfo{series}{Lect.\ Notes in Comput.\ Sci.}} \bibinfo{volume}{9035},
  \bibinfo{publisher}{Springer}, pp. \bibinfo{pages}{353--367},
  \doi{10.1007/978-3-662-46681-0\_28}.

\bibitemdeclare{phdthesis}{alglave}
\bibitem{alglave}
\bibinfo{author}{J.~\surnamestart Alglave\surnameend} (\bibinfo{year}{2010}):
  \emph{\bibinfo{title}{A Shared Memory Poetics}}.
\newblock Ph.D. thesis, \bibinfo{school}{Universit{\'e} Paris 7}.
\newblock \urlprefix\url{http://www0.cs.ucl.ac.uk/staff/J.Alglave/these.pdf}.

\bibitemdeclare{inproceedings}{boudol12}
\bibitem{boudol12}
\bibinfo{author}{G.~\surnamestart Boudol\surnameend},
  \bibinfo{author}{G.~\surnamestart Petri\surnameend} \&
  \bibinfo{author}{Serpette \surnamestart G.\surnameend}
  (\bibinfo{year}{2012}): \emph{\bibinfo{title}{Relaxed Operational Semantics
  of Concurrent Programming Languages}}.
\newblock In \bibinfo{editor}{B.~\surnamestart Luttik\surnameend} \&
  \bibinfo{editor}{M.~A. \surnamestart Reniers\surnameend}, editors: {\sl
  \bibinfo{booktitle}{Proc.\ of Combined 19th Wksh.\ on Expressiveness in Concurrency
  and 9th Wksh.\ on Structural Operational Semantics, EXPRESS/SOS 2012}}, {\sl
  \bibinfo{series}{Electron.\ Proc.\ in Theor.\ Comput.\
  Sci.}}~\bibinfo{volume}{89}, pp. \bibinfo{pages}{19--33},
  \doi{10.4204/eptcs.89.3}.

\bibitemdeclare{book}{foata}
\bibitem{foata}
\bibinfo{author}{P.~\surnamestart Cartier\surnameend} \&
  \bibinfo{author}{D.~\surnamestart Foata\surnameend} (\bibinfo{year}{1969}):
  \emph{\bibinfo{title}{Problemes combinatoires de commutation et
  r{\'e}arrangements}}.
\newblock {\sl \bibinfo{series}{Lect.\ Notes in Math.}}~\bibinfo{volume}{85},
  \bibinfo{publisher}{Springer}, \doi{10.1007/bfb0079468}.

\bibitemdeclare{book}{godefroid}
\bibitem{godefroid}
\bibinfo{author}{P.~\surnamestart Godefroid\surnameend} (\bibinfo{year}{1996}):
  \emph{\bibinfo{title}{Partial-Order Methods for the Verification of
  Concurrent Systems: An Approach to the State-Explosion Problem}}.
\newblock \bibinfo{publisher}{Springer}, \doi{10.1007/3-540-60761-7}.

\bibitemdeclare{article}{lamport}
\bibitem{lamport}
\bibinfo{author}{L.~\surnamestart Lamport\surnameend} (\bibinfo{year}{1979}):
  \emph{\bibinfo{title}{How to Make a Multiprocessor Computer That Correctly
  Executes Multiprocess Programs}}.
\newblock {\sl \bibinfo{journal}{IEEE Trans. on Comput.}}
  \bibinfo{volume}{28}(\bibinfo{number}{9}), pp. \bibinfo{pages}{690--691},
  \doi{10.1109/tc.1979.1675439}.

\bibitemdeclare{article}{mazurkiewicz}
\bibitem{mazurkiewicz}
\bibinfo{author}{A.~\surnamestart Mazurkiewicz\surnameend}
  (\bibinfo{year}{1995}): \emph{\bibinfo{title}{Introduction to Trace Theory}}.
\newblock {\sl \bibinfo{journal}{The Book of Traces}}, pp.
  \bibinfo{pages}{3--41}, \doi{10.1142/9789814261456_0001}.

\bibitemdeclare{inproceedings}{x86-tso}
\bibitem{x86-tso}
\bibinfo{author}{S.~\surnamestart Owens\surnameend},
  \bibinfo{author}{S.~\surnamestart Sarkar\surnameend} \&
  \bibinfo{author}{P.~\surnamestart Sewell\surnameend} (\bibinfo{year}{2009}):
  \emph{\bibinfo{title}{A Better x86 Memory Model: {x86-TSO}}}.
\newblock In \bibinfo{editor}{S.~\surnamestart Berghofer\surnameend},
  \bibinfo{editor}{T.~\surnamestart Nipkow\surnameend},
  \bibinfo{editor}{C.~\surnamestart Urban\surnameend} \&
  \bibinfo{editor}{M.~\surnamestart Wenzel\surnameend}, editors: {\sl
  \bibinfo{booktitle}{Proc.\ of 22nd Int.\ Conf.\ on Theorem Proving in Higher
  Order Logics, TPHOLs~2009}}, {\sl \bibinfo{series}{Lect.\ Notes in Comput.\
  Sci.}} \bibinfo{volume}{5674}, \bibinfo{publisher}{Springer}, pp.
  \bibinfo{pages}{391--407}, \doi{10.1007/978-3-642-03359-9\_27}.

\bibitemdeclare{inproceedings}{park}
\bibitem{park}
\bibinfo{author}{S.~\surnamestart Park\surnameend} \& \bibinfo{author}{D.~L.
  \surnamestart Dill\surnameend} (\bibinfo{year}{1995}):
  \emph{\bibinfo{title}{An Executable Specification, Analyzer and Verifier for
  {RMO} (Relaxed Memory Order)}}.
\newblock In: {\sl \bibinfo{booktitle}{Proc.\ of 7th Ann.\ ACM Symp.\ on
  Parallel Algorithms and Architectures, SPAA~'95}}, \bibinfo{publisher}{ACM},
  pp. \bibinfo{pages}{34--41}, \doi{10.1145/215399.215413}.

\bibitemdeclare{inproceedings}{sassone}
\bibitem{sassone}
\bibinfo{author}{V.~\surnamestart Sassone\surnameend},
  \bibinfo{author}{M.~\surnamestart Nielsen\surnameend} \&
  \bibinfo{author}{G.~\surnamestart Winskel\surnameend} (\bibinfo{year}{1993}):
  \emph{\bibinfo{title}{Deterministic Behavioural Models for Concurrency}}.
\newblock In \bibinfo{editor}{A.~M. \surnamestart Borzyszkowski\surnameend} \&
  \bibinfo{editor}{S.~\surnamestart Sokolowski\surnameend}, editors: {\sl
  \bibinfo{booktitle}{Proc.\ of 18th Int.\ Symp.\ on Mathematical Foundations
  of Computer Science, MFCS~'93}}, {\sl \bibinfo{series}{Lect.\ Notes in
  Comput.\ Sci.}} \bibinfo{volume}{711}, \bibinfo{publisher}{Springer}, pp.
  \bibinfo{pages}{682--692}, \doi{10.1007/3-540-57182-5_59}.

\bibitemdeclare{book}{sparc1994}
\bibitem{sparc1994}
\bibinfo{author}{\surnamestart {SPARC International Inc.}\surnameend} \&
  \bibinfo{author}{David~L. \surnamestart Weaver\surnameend}
  (\bibinfo{year}{1994}): \emph{\bibinfo{title}{The {SPARC} Architecture
  Manual}}.
\newblock \bibinfo{publisher}{Prentice-Hall}.

\bibitemdeclare{inproceedings}{yang}
\bibitem{yang}
\bibinfo{author}{Y.~\surnamestart Yang\surnameend},
  \bibinfo{author}{G.~\surnamestart Gopalakrishnan\surnameend},
  \bibinfo{author}{G.~\surnamestart Lindstrom\surnameend} \&
  \bibinfo{author}{K.~\surnamestart Slind\surnameend} (\bibinfo{year}{2004}):
  \emph{\bibinfo{title}{{Nemos}: A Framework for Axiomatic and Executable
  Specifications of Memory Consistency Models}}.
\newblock In: {\sl \bibinfo{booktitle}{Proc.\ of 18th Int.\ Parallel and
  Distributed Processing Symposium, IPDPS 2004}}, \bibinfo{publisher}{IEEE},
  pp. \bibinfo{pages}{31--40}, \doi{10.1109/ipdps.2004.1302944}.

\bibitemdeclare{inproceedings}{zhang}
\bibitem{zhang}
\bibinfo{author}{N.~\surnamestart Zhang\surnameend},
  \bibinfo{author}{M.~\surnamestart Kusano\surnameend} \&
  \bibinfo{author}{C.~\surnamestart Wang\surnameend} (\bibinfo{year}{2015}):
  \emph{\bibinfo{title}{Dynamic Partial Order Reduction for Relaxed Memory
  Models}}.
\newblock In: {\sl \bibinfo{booktitle}{Proc.\ of 36th ACM SIGPLAN Conf.\ on
  Principles of Language Design and Implementation, PLDI 2015}},
  \bibinfo{publisher}{ACM}, pp. \bibinfo{pages}{250--259},
  \doi{10.1145/2737924.2737956}.

\end{thebibliography}

\appendix

\section{Semantic Rules} \label{rules}

\paragraph{Small steps of a processor}

{
\renewcommand{\arraystretch}{3}

\[
\begin{array}{cc}
\infer{[] \Isame e'}{
}
&
\infer{e:bklg \Isame e'}{
  e \Isame e'
  &
  bklg \Isame e' 
}
\end{array}
\]

\[
\begin{array}{c}
\infer{(act:prg, bklg, eid) \xrightarrow{(eid,\main,act)}
  (prg, (eid,\shdw,act):bklg, eid+1)}{
   \shadows(act)
   &
   bklg \Isame (eid,\main,act)
}
\\
\infer{(act:prg, bklg, eid) \xrightarrow{(eid,\main,act)} 
      (prg, bklg, eid+1)}{
   \neg \shadows(act)
   &
   bklg \Isame (eid,\main,act) 
}
\\
\infer{(prg, newer \ap (le : older), eid) \xrightarrow{le} 
      (prg, newer \ap older, eid)}{
   older \Isame le
}
\\
\infer{(prg_0+prg_1, bklg, eid) \xrightarrow{le} c}{
  (prg_i, bklg, eid) \xrightarrow{le} c    
}   
\end{array}
\]

}

\paragraph{Small steps of the system}

\[
\infer{c \xrightarrow{(pid,le)} c[pid \mapsto lc']}{
   c(pid) = lc
   & 
   lc \xrightarrow{le} lc'
}
\]

\paragraph{Executions}

\[
\begin{array}{cc}
\infer{c \xRightarrow{[]} c}{
  \forall pid.\, c(pid) = ([], [], \_)
}
&
\infer{c \xRightarrow{e : es} c'}{
  c \xrightarrow{e} c''
  &
  c'' \xRightarrow{es} c'
}
\end{array}
\]

\paragraph{Normal executions}

\[
\begin{array}{cc}
\infer{(pid, le) \I{ss} (pid, le')}{
 le \Isame le'
}
&
\infer{(pid, le) \I{ss} (pid', le')}{
 pid \neq pid'
 &
 le \Idiff{ss} le'
}
\end{array}
\]


{
\renewcommand{\arraystretch}{3}

\[
\begin{array}{c}
\infer{[e] \prec e'}{
  e \prec e'
}
\quad
\infer{s:e \prec e'}{
  e \prec e'
}
\qquad
\infer{[e] \I{ss} e'}{
  e \I{ss} e'
}
\quad
\infer{s:e \I{ss} e'}{
  s \I{ss} e' 
  &
  e \I{ss} e'
}
\end{array}
\]

\[
\begin{array}{c}
\infer{ss \seq []}{
}
\\
\infer{[] \seq e: es}{
  []:[e] \seq es
}
\quad
\infer{[]:s \seq e: es}{
  s \I{[]} e
  &  
  s \prec e
  &
  []:(s:e) \seq es
}
\\
\infer{ss:s \seq e: es}{
  \neg (s \I{ss} e)
  &
  ss:s:[e] \seq es
}
\quad
\infer{ss:s:s' \seq e: es}{
  \neg (s \I{ss} e)
  &
  s' \I{ss:s} e
  &  
  s' \prec e
  &
  ss:s:(s':e) \seq es
}
\end{array}
\]
}

\end{document}